\documentclass[twocolumhn]{emulateapj}
\usepackage{graphicx}
\usepackage{times}

\usepackage{colordvi}
\usepackage{epsfig}
\usepackage{appendix}

\def \aa {{\it {\AA}}}
\def \mg {{Mg~II}}
\def \mbh{\ensuremath{M_{\rm BH}}}
\def \edr{\ensuremath{R_{\rm Edd}}}

\def \wmg{\ensuremath{{\rm W}_{\rm Mg\;II}}}

\def \wc4{\ensuremath{{\rm W}_{\rm C\;IV}}}
\def \kms{\ensuremath{{\rm km\;s}^{-1}}}
\def \zem{\ensuremath{z_{em}}}
\def \zab{\ensuremath{z_{abs}}}
\def \ebv{\ensuremath{E(B-V)}}

\def \bet{\ensuremath{\beta}}
\def \sig{\ensuremath{\sigma}}
\def \imag{\ensuremath{i\;{\rm magnitude}}}

\shorttitle{Associated {Mg~II} II absorbers}
\shortauthors{Khare et al.}
\begin{document}
\title{The dust, nebular emission and dependence on QSO radio properties of
the associated Mg II absorption line systems}

\author{Pushpa Khare}\affil{CSIR Emeritus Scientist, IUCAA, Ganeshkhind, Pune,
411007, India}

\author{Daniel Vanden Berk}\affil{Physics Department, St. Vincent College,
Latrobe, PA 15650, USA}

\author{Hadi Rahmani}\affil{School of Astronomy, Institute for Research in
Fundamental Sciences (IPM), PO Box 19395-5531, Tehran, Iran}

\author{Donald G. York}\affil{Department of Astronomy and Astrophysics,
University of Chicago, Chicago, IL 60637; Enrico Fermi Institute, University of
Chicago, Chicago, IL 60637}

\begin{abstract} We studied dust reddening and [O~II] emission in 
1730 \mg~ associated absorption systems (AAS; relative velocity with respect to
QSOs, $\le$ 3000 km s$^{-1}$; in units of velocity of light, \bet, $\le$ 0.01)
with 0.4 $\le z_{abs}\le$ 2 in the SDSS DR7, focusing on their dependence on
the radio and other QSO properties. We used control samples, several with
matching radio properties to show (i) AAS in radio detected (RD) QSOs
cause 2.6$\pm$0.2 times higher dust extinction than those in radio undetected
(RUD) ones which, in turn, cause 2.9$\pm$0.7 times the dust extinction in the
intervening systems; (ii) AAS in core-dominated QSOs cause 2.0$\pm$0.1
times higher dust extinction than in lobe-dominated QSOs; (iii) occurrence
of AAS is 2.1$\pm$0.2 times more likely in RD QSOs than in RUD QSOs and
1.8$\pm$0.1 time more likely in QSOs having black holes with masses larger than
1.23$\times 10^{9}$ M$_\odot$ than in those with lower mass black holes; (iv)
there is excess flux in [O~II]$\lambda$3727 emission in the composite spectra
of the AAS samples compared to those of the control samples, which is at the
emission redshift.  Presence of AAS enhances the O II emission from the AGN
and/or the host galaxy. This excess is similar for both RD and RUD samples, and
is 2.5$\pm$0.4 times higher in lobe-dominated compared to core-dominated
samples. The excess depends on the black hole mass and Eddington ratio. All
these point to the intrinsic nature of the AAS except for the systems with
$z_{abs}> z_{em}$ which could be infalling galaxies. 
\end{abstract}
\section{Introduction} The origin of associated absorption systems (AAS) in the
spectra of QSOs with \bet~$<$ 0.01\footnote{\bet~ is the relative velocity of
the absorption systems with respect to the systemic redshift of the QSO in
units of velocity of light} (the relative velocity of the AAS with respect to the QSO, hereafter V, $<$ 3000 km s$^{-1}$) is not very
well understood. Possibilities for their origin include (i) interstellar/halo
clouds in the host galaxy (e.g.  \citealt{Ch07}), (ii) material in the
core of the active galactic nucleus (AGN), within 10 pc of the black hole
\citep{BS97}, (iii) material within 30 kpc of the AGN, accelerated
by starburst shocks from the inner galaxy \citep[e.g.][]{FS07} and (iv)
clouds in galaxies clustered around the QSO \citep[e.g.][]{W08}. The study
of properties of a large sample of such systems and in particular, their
correlation with radio and other properties of the parent QSO, should provide
clues towards discerning between various scenarios. 

Several studies of the dependence of the properties of the AAS on QSO radio
properties have been undertaken in the past. Some studies found the frequency
of occurrence of AAS to depend on the radio properties of the QSO 
\citep[e.g.][]{An87,Fo88,Gan01,Ba02}
Other studies failed to find such dependence \citep[e.g.][]{Ve03} but
concluded that differences in the results could probably be attributed to
various differences in selection of the relatively small samples (50-100
systems). An excess of AAS in radio loud QSOs was also found by \citet{W08} 
in a large sample of SDSS QSOs.   

In a previous study \citep[][hereafter V08]{V08}, based on Sloan
Digital Sky Survey (SDSS) data release 3 (DR3), we had studied the average dust
extinction and average abundances of a homogeneous sample of 407 AAS (with
\bet~ $<$ 0.01; V $<$ 3000 km s$^{-1}$) using the method of composite spectra
(York et al. 2006; hereafter Y06).  Definite evidence of dust in the AAS was
obtained by comparing the composite spectra of the absorber sample with that of
a non-absorber (control) sample matching in \zem~ and \imag~ on a one to one
basis. The dust was found to be of SMC type with no evidence for the presence
of a 2175 \aa~ absorption feature.  The amount of dust extinction and the
frequency of occurrence of AAS in the sample were found to depend on radio
properties of the QSOs. 

A much larger (by a factor of $>$ 4) sample of \mg~ systems is now available
from the SDSS DR7 (Shen \& Menard 2012; hereafter SM12). With this increase in
size, it should be possible to gain further understanding of the associated
absorbers. In particular, we can study the dependence of their dust extinction
and star formation rate (SFR) (as measured from the [O~II]$\lambda$3727
emission line flux), on radio and other properties e.g. the black hole
mass (\mbh, determined from the widths of various emission lines) and Eddington
ratio (\edr~ which is the ratio of bolometric luminosity of the object to its
Eddington luminosity) of the QSOs. Black hole mass may be indicative of its
age.

In a merger-driven model of AGNs, supermassive black holes evolve through major
mergers which give rise to starbursts and also to accretion onto the nuclear
black hole (Sanders et al. 1988; Hopkins et al. 2005, 2006). In such models
large amounts of gas and dust are funneled inward which fuels the black hole
and also causes the obscuration of the young QSO. The dust might be cleared
during a transitional phase resulting in the emergence of a luminous blue QSO.
In this picture, reddening is correlated with the evolutionary stage of
the QSO.  Recently, Shen \& Menard (2012) have shown that QSOs with associated
absorbers with \bet~$<$ 0.005 (V $<$ 1500 km s$^{-1}$), exhibit enhanced star
formation.  They suggest that these absorbers could be large-scale outflows
indicative of the transitional phase in a merger-driven evolutionary scenario
for QSOs. Based on the smaller dust extinction and SFR, they conclude that the
systems with \bet~$>$ 0.005 (V $>$ 1500 km s$^{-1}$) originate in intervening
absorbers.  

In this paper we present the results of our study of the SM12 sample of AAS,
using the method of composite spectra. We particularly focus on the dependence
of AAS properties on the radio and other properties of the QSOs. The
details of the sample and sub-samples thereof as well as the method of analysis
are presented in section 2, results are presented in section 3 and conclusions
are presented in section 4.

\section{Sample selection and analysis} 
\subsection{Main sample and sub-samples} As mentioned above we used the sample
compiled by SM12 (their Table 1) from the SDSS DR7. Their sample contains 1937
systems in non-BAL QSOs with \bet~$<0.01$ (V $<$ 3000 km s$^{-1}$), spanning
the redshift range of 0.4-2 and with rest equivalent width \wmg~ ranging from
0.22 to 6.8 \aa.  Five of these have \wmg~ $<$ 0.3 \aa. There are 92
sight-lines having two AAS each and six sight-lines having three AAS each.  For
this study, we focused primarily on QSOs having only one AAS along their lines
of sight, although some statistics for QSOs with multiple AAS are also
presented.  Also, we restricted our sample to \wmg $>$ 0.3 \aa~ so that our
results can be compared with those of V08 and Y06. Our full sample, S1, thus
consists of 1730 AAS having \wmg $\ge$ 0.3 \aa. 

We compiled several sub-samples from S1 by dividing it roughly in half, based
on various QSO and absorber properties: \wmg, \imag, radio properties, \mbh,
and \edr. Among the radio properties, we consider whether the QSOs have
been radio detected (RD) or undetected (RUD) in the FIRST survey, as well as
whether the sources are core dominated (CD) or lobe dominated (LD). Among our
sample of 1730 AAS, 263 have been detected while 1341 have been undetected in
the FIRST survey. 67 of the 263 RD QSOs, 67 are lobe dominated while 196 are
core dominated. The division of S1 based on \bet~ was done using the criterion
of SM12 (\bet~$<$0 (V $<$ 0 km s$^{-1}$), 0 $\le$ \bet~ $<$ 0.005 (0 $\le$ V
$<$ 1500 km s$^{-1}$) and \bet~$\ge$ 0.005 (V $\ge$ 1500 km s$^{-1}$) so that
our results can be compared with their conclusions.  The radio and other
properties (e.g. \mbh~ and \edr) of QSOs used for defining the sub-samples were
taken from Shen et al.  (2011). Throughout this study, we use the emission
redshifts as given by Hewett \& Wild (2010) and use the relative velocities of
AAS with respect to these as given by SM12.  Details of the sub-samples are
given in Table 1 which lists number of systems and the average values of \wmg,
\zab, \bet, \imag, \mbh (this is given in units of M$_\odot$ throughout the
paper), and \edr. 

\subsection{Composite spectrum} The method of forming composite spectra is
described in detail by Y06 and V08.  We describe it briefly here. First, the
spectra of individual QSOs, corrected for Galactic reddening, were shifted to
the absorber/QSO rest-frame and resampled onto a common pixel-to-wavelength
scale. Pixels flagged by the spectroscopic pipeline as possibly bad in some way
(Stoughton et al. 2002) were masked and not used in constructing the
composites. Also masked were the pixels within 5 \aa~ of the expected line
positions of detected intervening absorption systems unrelated to the target
system. The geometric mean of all contributing spectra was then calculated for
each pixel.  The median/mean composite for [O~II] emission line studies was
obtained by first fitting the continuum to $\sim$ 30 \aa~ wide regions around
3727 \aa~ in the QSO/absorber rest-frame and then resampling the continuum
subtracted spectra to a common wavelength scale.

We calculate the \ebv~ values for various samples by comparing the composite
spectrum of each sample with that of the corresponding control samples (sample
of QSOs not having AAS, matching one to one in \zem~ and \imag~ with the QSOs
in the absorber sample), and fitting an SMC curve (Pei 1992) to the extinction
curve so obtained. In deriving this extinction curve, we have normalized the
two composites at 3000 \aa, which was the value used by SM12. However, we have
verified that normalizing at longer wavelengths does not affect the \ebv~
values. The values are also independent of whether QSO rest-frame or absorber
rest-frame composites are used.  To construct the absorber rest-frame composite
of the control sample, the spectrum of each QSO in the sample was shifted to
the rest-frame of the absorber towards the corresponding QSO (matching in
\imag~ and \zem) in the absorber sample. The typical 1 \sig~ errors in the
derived \ebv~ values are generally smaller than 0.003 (The errors of the
relative flux density values are calculated using the variance formula for the
propagation of errors of a geometric mean. see Y06 and V08 for a detailed
analysis of errors). The Milky Way extinction curve does not fit well for any
of the samples, and the dust seems to be of SMC type, with no evidence of the
2175 \aa~ bump. The \ebv~ values so determined are given in the third column of
Table 2. We note that any difference in the \ebv~ values of different
sub-samples indicates (i) a difference in the dust column density caused either
by different gas column density or by different dust-to-gas ratio, or (ii)
different properties of the dust. In QSO absorbers, the extinction curve is
found to be same (SMC type) for all the samples. Thus, the dust properties
appear to be similar. The dust content is known to be correlated with \wmg~
(Y06, V08, Wild et al. 2006, Menard et al. 2011). Thus, the dust column
may be correlated with \wmg~ and in case we find the values of \ebv~ to be
different for sub-samples having similar \wmg, then it may indicate a
difference in the dust-to-gas ratio for the two sub-samples.  
\subsection{Control samples} For sub-samples S1-S8 and S13-S16, we
used the control samples which were used by SM12 (and kindly provided by them).
In the construction of these samples the radio properties of the QSOs were not
taken into account which is also the case for these sub-samples. However, it is
known that the RD QSOs are intrinsically redder than the RUD QSOs (e.g.\ see
Figure 6 of Kimball et al. 2011; hereafter K11) irrespective of
presence/absence of AAS.  The geometric mean composites for the samples of 4714
RD and 65253 RUD DR7 QSOs in K11 were kindly provided to us by the authors.
Fitting an SMC extinction curve to the ratio of the two composites, we estimate
the relative \ebv~ between the two samples to be 0.042. Another selection
effect could be important: the SDSS QSO target selection is made on the basis
of (blue) color. However a number of QSOs which may not satisfy the color
selection criteria are also observed based on other criteria, mainly, their
luminosity in other bands like the radio and hence need not have typical QSO
colors and could be reddened by selection. About 70 QSOs in both S9 and S10 are
not color selected, i.e. a much higher fraction of QSOs in the RD sample are
not color selected and hence could be reddened. In addition, the intrinsic
reddening in RD QSOs could also depend on their radio morphology. On the basis
of a complete sample of 4714 SDSS QSOs in DR7 with determined radio properties
and with FIRST flux S20 $>$ 2 mJy, K11 found that radio sources with unresolved
cores have higher reddening.  It is therefore necessary to construct control
samples also based on their radio properties for sub-samples S9-S12.  

We constructed such control samples for sub-samples S9 and S10 (matching one to
one in \zem~ and \imag~) from the RD and RUD QSOs (respectively) in SDSS DR7
which do not have AAS in their spectra.  The \ebv~ values obtained using these
control samples should be independent of any intrinsic reddening in the QSOs
caused by their radio properties and should reflect the reddening caused by the
presence of AAS. As the control sample now has the matching radio properties,
non-color selection should be equally probable in the sub-sample and its
control sample.  We also constructed control samples of lobe dominated and
core dominated QSOs from such QSOs in DR7 without AAS for S11 and S12
respectively. The values of E(B-V) for all these sub-samples (S9-S12), given in
Table 2 are calculated using these control samples with matching radio
properties in addition to matching \zem~ and \imag. 
\section{Results} 
\subsection{Dependence of the frequency of occurrence of AAS on the radio
and other properties of QSOs} Among DR7 non-BAL QSOs with redshifts between 0.4
and 2.0, 68755 QSOs have been observed by the FIRST radio survey (Becker,
White, and Helfand 1995).  Out of these, 6366 are radio detected, 4668 are
core-dominated, and 1698 are lobe-dominated.  Thus, the fraction of RD QSOs
among all QSOs in DR7 in the relevant redshift range is 0.093$\pm$0.001.  Out
of 1730 QSOs in our sample (having a single AAS each), 1604 have been observed
by FIRST: 263 are RD, 196 are core-dominated, and 67 are lobe-dominated.
Additionally, out of the 98 QSOs in the SM12 sample having multiple AAS, 90
have been observed by FIRST: 30 are RD (26 being CD) while 60 are RUD QSOs. The
fraction of RD QSOs among the QSOs having AAS (single or multiple) is
0.17$\pm$0.01. The occurrence of AAS is higher by a factor of 2.1$\pm$0.1 in RD
QSOs compared to RUD QSOs.  Similarly, while a fraction $\sim$0.11$\pm$0.02 of
RD QSOs in our sample have multiple AAS, the fraction is $\sim$0.045$\pm$0.006
for RUD QSOs in this sample: the incidence of multiple AAS is $\sim2.5\pm0.6$
times as likely in RD QSOs compared to the RUD ones.  These values remain
unchanged (to within 1 \sig) if the sample is restricted to the systems with
\bet~ $<$ 0.005 (V $<$ 1500 km s$^{-1}$, which are intrinsic systems according
the results of SM12.  Thus, the dependence of incidence of AAS on radio
properties is the same for all systems with \bet~ $<$0.01. The occurrence of
AAS is significantly related to the radio properties of the QSOs which is
consistent with the results of most previous studies mentioned in section 1.

The fraction of core-dominated QSOs among the RD DR7 QSOs is
$\sim$0.73$\pm$0.01 while in our sample, selected by the presence of (single
or multiple) AAS, it is $\sim$0.76$\pm$0.07. The presence of AAS seems to be
independent of the morphology of the radio source.  We caution that the numbers
here are small and the statistics may not be very meaningful.  These results
are unchanged if we restrict the analysis to systems with \bet~ $<$ 0.005.  

Among the 73181 non-BAL QSOs in the redshift range of 0.4 and 2.0 in DR7, 47521
QSOs have log(\mbh) $\le$ 9.1 while 25660 have log(\mbh) $>$ 9.1.  AAS are
present in a fraction of 0.028$\pm$0.001 of QSOs having smaller black hole mass
while for more massive black holes the values are 0.033$\pm$0.001. The values
are thus significantly different. The corresponding values for QSOs with
log(\edr)$\le$-0.81 and $>$-0.81 are 0.024$\pm$0.001 and 0.022$\pm$0.001
respectively. Thus, considering the black hole mass, QSOs with older
black holes have a higher frequency of occurrence of AAS.
No such difference exists for samples of lower and higher Eddington ratios.  

\subsection{Reddening} The reddening for S1 can be directly compared with
the reddening obtained for intervening systems by Y06 as both samples are from
the SDSS data and use the same method of composite spectra and as the selection
criterion for the two samples are identical except for the range of relative
velocities with respect to the QSOs. The \ebv~ for S1 is 3.2$\pm$0.8 times
higher than for the intervening systems for which the \ebv~ is 0.013. A
comparison with the results of V08 shows that the \ebv~ values obtained here
are larger by factors between 1.5-2, while the dependence of \ebv~ on \wmg,
\imag~ and radio properties is same. This could partly be because of the
higher fraction (15.2\%)of RD QSOs in our sample as compared to (9.9\%) in the
sample of V08 and partly a small sample effect.  

As seen from Table 2, the reddening is higher in fainter QSOs (S3) compared to
brighter ones (S2). This could be due to the higher average \wmg~ (1.61 \aa) in
S3 compared to that (1.32 \aa) in S2 as reddening is sensitive to the \wmg~
values (see \ebv~ values for S4 and S5 in Table 2).  Reddening in AAS with
\bet~ $>$ 0.005 (S8) is significantly smaller than the values for samples with
smaller \bet~ (S6 and S7) as found by SM12, but is still significantly higher
by a factor of 2.1$\pm0.5$ than that in intervening systems. The sub-sample
with higher \mbh~ (S14) is brighter (see Table 1) than that with lower \mbh~
(S13), while the \ebv~ values are equal for both sub-samples.  The sub-sample
with lower \edr~ (S15) is more reddened compared with the sub-sample having
higher Eddington ratio (S16).  
\subsubsection{Dependence of reddening on radio properties of the QSOs} It
is clear from the \ebv~ values that the AAS in RD QSOs are indeed dustier (by a
factor of 2.6$\pm$0.2) than those in RUD QSOs. Among the average properties
(See Table 1) of RD and RUD QSOs (sub-samples S9 and S10), the RD QSOs have
only marginally higher \wmg, $i$ band brightness, and \edr. The excess
extinction in the RD sub-sample over that in the RUD QSOs thus, can not be
possibly be accounted for by the differences in these average properties.  As
the difference in reddening in RD QSOs can not be accounted by the difference
in \wmg~ therefore, as noted in section 2.2, the AAS towards RD QSOs may have
higher dust-to-gas ratio.  Differences can also be seen among the core and
lobe-dominated sub-samples, in that core-dominated QSOs are 2.0$\pm0.1$ times
more reddened.  Both classes of RD QSOs (S11 and S12) are significantly more
reddened compared to the RUD QSOs. This is consistent with the results of V08
but is in contrast to the results of K11 who found that only radio sources with
unresolved cores have higher reddening, and other classes of RD QSOs have
reddening comparable to that of the RUD QSOs. Also, the reddening in AAS in RUD
QSOs is 2.9$\pm$0.7 times higher than that in the intervening systems (Y06). 

To check if the reddening in RD QSOs with AAS is correlated with \wmg, we
divided sub-sample S9 into two roughly equal halves (S9a and S9b) depending on
\wmg. The results for these are given in Tables 2.  These confirm that the RD
QSOs with stronger \mg~ absorption lines are considerably more reddened
compared to those having weaker \mg~ lines; the difference between the \ebv~
values of the two samples being 0.093.  The average \wmg~ for the two
sub-samples S9a and S9b are 2.5 and 0.9 \aa, which are very similar to those of
sub-samples S5 and S4 (2.1 and 0.83 \aa) respectively, for which the \ebv~
differ by only 0.029.  Thus RD QSOs have a stronger dependence of reddening on
\wmg. We also divided the RUD sub-sample (S10) into two halves (S10a and S10b)
depending on \wmg. The results for these are given in the last rows of Tables
2.  The average \wmg~ values for sub-samples S10a and S10b and the difference
in the \ebv~ values for these samples are similar to that between
the values for sub-samples S5 and S4. 

The main conclusion from this study is that the AAS in RD QSOs are
significantly more reddened than those in RUD QSOs which, in turn, are
significantly more reddened than the intervening systems.

\begin{deluxetable*}{lcrllrlll}
\tablecaption{Definitions and properties of sub-samples of AAS.}
\tabletypesize{\scriptsize}
\tablehead{
\colhead{\bf Sample}& \colhead{\bf Selection }&\colhead{\bf No.
of}&\colhead{\bf $<$\wmg$>$}&\colhead{$\bf <$z$_{em}>$}&\colhead{ $\bf
<\beta>$}&\colhead{ $\bf <$m$_i^a>$}&\colhead{\bf Log($<$M$^b_{\rm
BH}>$)}&\colhead{\bf Log($<$R$_{\rm Edd}>$)}\\ {\bf number}&{\bf criterion
}&\colhead{\bf systems}&\colhead{\bf in \aa}&&x10$^{3}$&&&}
S0&Sample of SM12&1937&1.43&1.27&1.3&18.56&9.04&-0.64\\
S1&Full sample&1730&1.46&1.28&2.16&18.59&9.04&-0.63\\
S2&m$_{i \le }$18.68&862&1.32&1.26&2.46&18.11&9.15&-0.58\\
S3&m$_{i> }$18.68&868&1.61&1.30&1.85&19.08&8.93&-0.69\\
S4&W$_{\rm Mg\;II}\le$1.24 \aa  &866&0.83&1.26&2.27&18.46&9.06&-0.64\\
S5&W$_{\rm Mg\;II}>$1.24 \aa  &864&2.10&1.30&2.04&18.73&9.01&-0.62\\
S6&\bet$ <$ 0.0&510&1.49&1.29&-1.39&18.62&9.05&-0.67\\
S7&0.0$\le$\bet $<$ 0.005&841&1.47&1.22&1.83&18.64&9.04&-0.70\\
S8&\bet$\ge$ 0.005&379&1.42&1.40&7.66&18.46&9.01&-0.47\\
S9&Radio-detected (RD)&263&1.67&1.18&1.79&18.45&9.03&-0.54\\
S10&Radio-undetected (RUD)&1341&1.43&1.30&2.25&18.62&9.04&-0.65\\
S11&Lobe-dominated (LD)&67&1.46&1.18&1.64&18.45&9.11&-0.42\\
S12&Core-dominated (CD)&196&1.74&1.18&1.85&18.45&9.01&-0.58\\
S13&Log(M$_{\rm BH}$)$\le9.09$&874&1.47&1.12&2.23&18.72&8.66&-0.49\\
S14&Log(M$_{\rm BH}$)$>9.09$&856&1.46&1.44&2.08&18.47&9.42&-0.85\\
S15&Log(R$_{\rm Edd}$)$\le$-0.81&870&1.55&1.20&1.59&18.78&9.19&-1.09\\
S16&Log(R$_{\rm Edd}$)$>$-0.81&860&1.38&1.36&2.73&18.41&8.88&-0.411\\
\tableline
\multicolumn{9}{l}{ a: SDSS $i$ magnitude, corrected for Galactic extinction.}\\
\multicolumn{9}{l}{ b: In units of M$_\odot$}\\
\end{deluxetable*}
\begin{deluxetable*}{cccc}
\tabletypesize{\scriptsize}
\tablecaption{$E(B-V)$ and [O~II]$\lambda$3727 flux excess for sub-samples of
AAS.}
\vspace*{0.1in}
\tablehead{
\colhead{\bf Sample}&\colhead{\bf Selection
}&\colhead{\bf \ebv$^a$ w.r.t }
&\colhead{\bf [O~II] flux w.r.t}\\
{\bf number}&{\bf criterion}&
{\bf control samples}&{\bf control samples$^b$}}
S1&Full sample&0.041&1.92$\pm0.04$\\
S2&m$_{i \le }$18.68&0.031&1.84$\pm0.04$\\
S3&m$_{i> }$18.68&0.049&2.07$\pm0.05$\\
S4&W$_{\rm Mg\;II}\le$1.24 \aa  &0.026&1.9$\pm0.05$\\
S5&W$_{\rm Mg\;II}>$1.24 \aa  &0.05-&1.94$\pm0.04$\\
S6&\bet$ <$ 0.0&0.047&1.32$\pm0.05$\\
S7&0.0$\le$\bet $<$ 0.005&0.055&2.32$\pm0.06$\\
S8&\bet$\ge$ 0.005&0.027&1.52$\pm0.07$\\
S9&Radio-detected (RD)&0.097&1.73$\pm0.04$\\
S10&Radio-undetected (RUD)&0.038&1.87$\pm$0.06\\
S11&Lobe-dominated (LD)&0.056&2.08$\pm$0.13\\
S12&Core-dominated (CD)&0.109&1.44$\pm$0.04\\
S13&Log(M$_{\rm BH}$)$\le9.09$&0.038&1.76$\pm0.04$\\
S14&Log(M$_{\rm BH}$)$>9.09$&0.041&2.40$\pm0.08$\\
S15&Log(R$_{\rm Edd}$)$\le$-0.81&0.058&2.19$\pm0.05$\\
S16&Log(R$_{\rm Edd}$)$>$-0.81&0.025&1.55$\pm0.04$\\
S9a$^c$&RD,W$_{\rm Mg\;II}>$1.4&0.149&2.10$\pm$0.08$$\\
S9b$^d$&RD,W$_{\rm Mg\;II}\le$1.4&0.056&1.44$\pm$0.04$$\\
S10a$^e$&RUD,W$_{\rm Mg\;II}>$1.23&0.057&2.38$\pm$0.13$$\\
S10b$^f$&RUD,W$_{\rm Mg\;II}\le$1.23&0.018&1.48$\pm$0.06$$\\
\tableline
\multicolumn{4}{l}{a: One \sig~ errors in \ebv~ are typically smaller than
0.003.}\\
\multicolumn{4}{l}{b: Ratio of flux in [O II] emission line in the sample composite to that in the composite of the }\\
\multicolumn{4}{l}{corresponding control sample. Note that the control samples used for S9, S10, s11, S12 and }\\
\multicolumn{4}{l}{for S9a, S9b, S10a and S10b have matching radio properties.}\\
\multicolumn{4}{l}{c: S9a: 129 systems belonging to sub-sample S9 and having \wmg
$>$ 1.4 \aa.}\\
\multicolumn{4}{l}{d: S9b: 134 systems belonging to sub-sample S9 and having \wmg
$\le$ 1.4 \aa.}\\
\multicolumn{4}{l}{e: S10a: 666 systems belonging to sub-sample S10 and having \wmg
$>$ 1.23 \aa.}\\
\multicolumn{4}{l}{f: S10b: 675 systems belonging to sub-sample S10 and having \wmg
$\le$ 1.23 \aa.}\\
\end{deluxetable*}

\subsection{Emission lines from host galaxies/absorbers}
\subsubsection{Excess emission flux in QSOs with AAS} 
SM12 found excess emission flux in the [O~II]$\lambda$3727 line (hereafter
EEFOII) in the composite spectrum of QSOs having AAS with \bet~$<$\,0.005 over
that in the composite spectra of the control sample of QSOs without AAS both
made at the emission rest frame. These results were interpreted as indicating
that the EEFOII originates in the host galaxies of QSOs having AAS with
\bet~$<$\,0.005, and that it is a measure of SFR in those galaxies.  We discuss
the [O~II] emission further below. 
 
In order to study the emission lines, we constructed continuum-subtracted
composites (median and mean) of various sub-samples as well as those of the
corresponding control samples in the QSO rest-frame. We also constructed
corresponding composites in the absorber rest-frames. For this, the spectrum of
each QSO in the control sample was shifted to the rest-frame of the absorber
towards the corresponding QSO (matching in \imag~ and \zem) in the absorber
sample.  We observed that the median composite is not very meaningful when
constructed in the absorber rest-frame as the peak of the [O~II]$\lambda$3727
emission line which gets contribution from the AGN and its host galaxy
at the emission redshift) is shifted by different amounts (depending on the
value of \bet~) and as a result the total flux in the line in the median
composite is not the same as that in the median composites in the emission
rest-frame. This problem does not appear if we construct the mean composites
(without clipping in the emission line region).  We have therefore, used the
mean composites for measuring the EEFOII.  The values of the ratio of the [O
II] flux in the composite spectrum of the  sample to that in the composite of
the corresponding control sample for various sub-samples using the control
samples described in section 2 are listed in column 4 of Table 2. In most cases
these are independent of whether the mean composites are made in the emission
or absorption rest-frames, we mention the exception to this below. 

It can be seen that a significant EEFOII is seen for all sub-samples.  The
magnitude of EEFOII depends on values of QSO brightness, \bet, black hole mass,
Eddington ratio, and on radio morphology. The EEFOII does not depend on the
strength of absorption lines (S4 and S5). This is different from what is seen
for the intervening systems where the [O~II] emission flux is proportional to
\wmg~ (Noterdaeme et al. 2010; Menard et al.  2011). As will be seen below,
this is due to the different fractions of RD and RUD QSOs in these samples. The
sub-samples with lower \mbh~ and higher \edr~ have lower EEFOII.  Finally, the
EEFOII in the S7 sample (0$\le$ \bet $<$0.005) is much higher than that in S6
(\bet $<0$) and S8 (\bet $\ge$0.005). It thus appears that the hosts of AAS
with \bet~ between 0 and 0.005 are different from those with \bet~ $<$ 0 and
$>$ 0.005. 

\begin{figure}
\epsscale{1.0}
\plotone{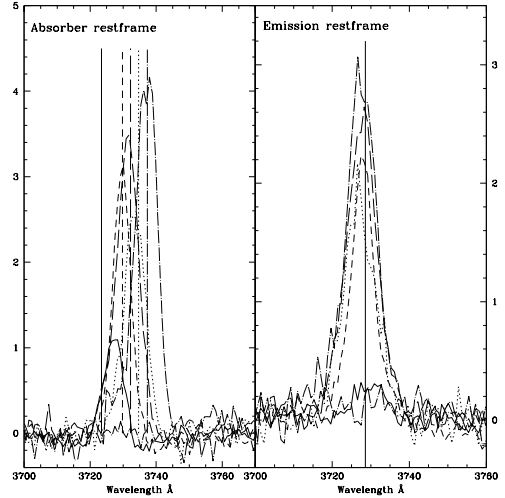}
\caption{Left panel shows the absorber frame residual spectra (difference
between the absorber rest-frame composites of the absorber sample and those of
corresponding control sample) for various \bet~ dependent samples. The samples
plotted are: S6 (solid line), $0 < v < 200 \,\kms (\bet~=0-0.0007)$
(short dashed line), $200$\,\kms\,$< v < 400$\,\kms  
(\bet~=0.0007-0.0013) (long dashed line), $400$\,\kms\,$< v < 600$\,\kms 
(\bet~=0.0013-0.002) (dotted line) and $600$\,\kms\,$< v < 800$\,\kms 
(\bet~=0.002-0.0027) (dash-dotted line).  The vertical lines indicate the
effective wavelengths of the [O~II] doublet (3728.6 \aa~ which is the value
used by Hewett \& Wild) at the mean values of the relative velocities for each
sample. The right panel shows the same but for emission rest-frame
composites.\label{fig1}}
\end{figure}

In order to confirm that the EEFOII arises in the emission rest-frame, we have
plotted in the left panel of Figure 1, the excess flux in the [O~II] line in
the absorption rest-frame composite spectra of various \bet~ dependent
sub-samples over that in the composite of the corresponding control samples.
For this figure we constructed additional sub-samples having relative velocity
between (i) 0 and 200 \kms (\bet~=0-0.0007), (ii) 200 and 400 \kms 
(\bet~=0.0007-0.0013), (iii) 400 and 600 \kms (\bet~=0.0013-0.002), and
(iv) 600 and 800 \kms (\bet~=0.002-0.0027). The number of systems in
these sub-samples are 214, 185, 110, and 106 respectively.  It is clear that
the line centers in the absorber rest-frame composites are progressively
shifted towards longer wavelengths with increasing \bet~ values of the
sub-samples. We have plotted the positions of the [O II] lines at the mean
velocities of the samples as vertical lines. For all samples with \bet~ $>$0,
these are roughly consistent with the centers of the corresponding emission
excess confirming that the [O II] emission is in the QSO rest-frame. For
\bet~$<$0 sample the line shows up at $\sim$3728 \aa~ in the absorber
rest-frame composite. The [O II] emission for these systems occurs at the
absorber rest-frame and does not show up in the emission rest-frame composite.
It thus seems possible that these absorbers are galaxies which are falling
towards the QSOs. In the right panel of Figure 1, we have plotted the
difference between the composites of the same samples and the composites of the
corresponding control samples in the emission rest-frame.  The excess is
centered around $\sim$3728\,\aa~ and clearly shows that the excess emission
occurs at the QSO redshift as was observed by SM12, and may originate in the
host galaxy of the individual QSOs. Note that the excess emission is not
seen here for the \bet~ $<$ 0 sample. 

We have measured the FWHMs of the excess [O~II] emission in the composites
(like those plotted in the right panel of Figure 1) of various sub-samples and
these values all fall between 8 and 10 \aa. To estimate the errors in FWHM
caused by the errors in \zem, we generated 100 composite spectra for each of
the samples plotted in Figure 1, each time choosing random \zem~ within the one
sigma values given by given by Hewett \& Wild (2010). These errors in \zem are
smaller than 0.006 for QSOs in our sample and in most cases are in fact smaller
than 0.0025. The one $\sigma$ error in FWHMs as measured from the FWHMs of the
of the 100 composites for each of the samples are all smaller than 0.5 \aa.
The FWHMs, thus, are large and are very similar to the FWHMs of the [O~II]
emission lines themselves in the spectra of the composites of the control
samples.  It therefore appears that the excess flux originates in the same
regions as the [O II] emission lines of QSOs.

\subsubsection{Dependence of the EEFOII on radio properties} Figure 6 of K11
shows that the flux of the [O~II]$\lambda$3727 line is significantly higher in
the geometric mean composite of all radio QSOs as compared to that for all RUD
QSOs in DR7. K11 have also studied the dependence of emission line fluxes in
the median composite spectra for several emission lines on the radio properties
of the QSOs. They find that the fluxes depend on the type of radio sources (see
their Figure 7). 

In our samples the values of EEFOII in RD QSOs are similar to that in RUD QSOs
(with AAS). The EEFOII for absorbers in the lobe-dominated QSOs (S11) however,
is higher than that for the core-dominated QSOs (S12). Thus, on the whole
the EEFOII in RD and RUD QSOs is same but in RD QSOs the EEFOII is dependent on
the radio morphology. It is interesting to note that in CD QSOs it is even
lower than that in the RUD QSOs. 

A comparison of the EEFOII for the RD sub-samples S9a and S9b defined on the
basis of \wmg, shows that among the RD QSOs, the EEFOII is much stronger in
QSOs having AAS with larger \wmg.  The same also holds if we divide the RUD
sub-sample S10 into two halves (S10a and S10b) depending on \wmg.  Thus, the
EEFOII depends on the strength of the \mg~ lines in the AAS in both RD and RUD
sub-samples. This dependence on \wmg~ is qualitatively similar to what has been
found for intervening systems (Noterdaeme et al. 2010; Menard et al. 2011).  

\section{Discussion} Using the largest sample of AAS compiled so far, we have
found that the incidence of {Mg~II} AAS significantly depends on the radio
properties of the QSOs. Radio detected QSOs are 2.1$\pm$0.2 times more likely
to have AAS as compared to RUD QSOs. Among the RD QSOs, the incidence of AAS
does not seem to depend on the morphology of the radio source. This is
inconsistent with earlier studies (mentioned in section 1) that found that the
incidence of AAS is either independent of the radio properties of the QSOs
(Vestergaard 2003) or it is higher in lobe-dominated QSOs (e.g. Aldcroft et al.
1994). We also find a significantly higher frequency of occurrence of AAS in
QSOs with higher black hole masses compared with that in QSOs with smaller
black hole masses.

It is clear from our results that the radio properties of the QSOs play an
important role in influencing the reddening properties of the AAS. The dust
extinction is higher in the AAS in RD QSOs than that in RUD QSOs by a factor of
2.6$\pm$0.2.  Among the RD QSOs, the CD QSOs are more reddened by a factor of
2.0$\pm$0.1 as compared to the LD QSOs. The reddening in the AAS in RUD QSOs is
also higher than that in the intervening systems (by a factor of 2.9$\pm0.7$).
For both types of QSOs (RD and RUD), the reddening in the AAS depends on the
strength of absorption lines; AAS with stronger lines show higher reddening.
Based on the dust extinction alone, the AAS in both RD and RUD QSOs appear to
be intrinsic to the QSO. The reddening is insensitive to the black hole mass
but depends on the Eddington ratio in the sense that AAS in 
QSOs having lower Eddington ratios have higher dust extinction.  

The dust extinction is found to be similar in AAS with \bet~ $<$0 and with
\bet~ between 0 and 0.005. The AAS with \bet~ $>$ 0.005 are found to have
smaller dust extinction by a factor of 1.7-2.0, however, the extinction is
higher than that in the intervening systems (Y06) by a factor of 2.1$\pm0.5$,
and therefore those AAS may not be intervening as suggested by SM12.

We note that there is always a selection effect acting against the very
dusty systems which will not be observable in a flux limited survey. This
however, applies both to the AAS as well as the intervening systems. Thus, the
difference in the dust content of these two types of systems as obtained by
comparing our results with those of Y06 is real as both are samples are taken
from the SDSS. We have earlier shown (Khare et al. 2007), from a study of
abundances in samples of DLAs and sub-DLAs observed at high resolution and also
the average abundances in SDSS Mg II systems at redshifts similar to what we
are studying here, that the obscuration bias is not likely to be important.
Either very dusty systems do not exist or there may be a bi-modal distribution
of dust and there may exist a population of completely dust obscured QSOs. Our
results do not apply to such a population.

We have clearly demonstrated that the excess emission in the [O~II] line
originates at the emission redshift of the QSOs (except for the \bet~$<$0
sample for which it occurs close to absorption redshift). This could be due to
star formation in the host galaxy. However, for all sub-samples, the FWHM of
the excess flux distribution is large ($\sim$ 8-10 \aa) and is similar to the
FWHM of the total [O~II] emission. Thus, the EEFOII may also originate in the
same regions in the parent AGN and the host galaxy which emit the [O II] line
and the presence of AAS enhances the O II emission from these regions.  

In the the whole sample, the EEFOII does not depend on the Mg~ II equivalent
width because of the different fractions of RD and RUD QSOs in the \wmg~
dependent sub-samples. This is clear from the fact that when we consider the RD
and RUD samples separately, the EEFOII is significantly higher for QSOs having
AAS with larger \wmg~, similar to what is observed for the intervening
absorbers. The values of the EEFOII are within the range of fluxes determined
for intervening Mg~ II systems (Noterdaeme et al. 2010). The EEFOII is similar
for the RD and RUD QSOs; however, among the RD QSOs, the EEFOII is higher for
the LD QSOs as compared to the CD QSOs.  Thus, the dust extinction and EEFOII
seem to be anti-correlated among the CD and LD QSOs, which is expected except
that the dust extinction seems to be too small to explain the difference in
EEFOIIs. The EEFOII does depend on the \mbh~ and \edr. EEFOII is higher in AAS
in QSOs having higher \mbh~ and having lower \edr.

\section{Conclusions}  
We have studied the properties of associated absorption systems in the redshift
range of 0.4 to 2.0, in the spectra of 1730 SDSS QSOs. The main conclusions are
as follows:

\begin{enumerate}
\item The average dust extinction is found to be of SMC type with no evidence
for the 2175 \aa~ bump.
\item The dust extinction is 3.2$\pm$0.8 times greater than in the intervening
\mg~ absorbers with similar selection criteria. 
\item By using the control samples comprised only of RD, RUD, lobe-dominated
and core-dominated QSOs, we find that (a) the AAS in RD QSOs are 2.6$\pm$0.2
times more dusty compared to the AAS in RUD QSOs; (b) the reddening due to
AAS in RD QSOs has a stronger dependence on \wmg~ compared to that in RUD QSOs;
(c) among the RD QSOs, the AAS in the core-dominated QSOs have 2.0$\pm$0.1 times
higher dust extinction compared to those in the lobe-dominated QSOs; (d) the
reddening in the AAS in RUD QSOs is 2.9$\pm$0.7 times higher than that in
intervening absorbers.
\item The reddening does not depend on the black hole mass and thus its
age.
\item The reddening does depend on the Eddington ratio, systems with
smaller \edr~ have higher reddening. \item The occurrence of AAS is 2.1$\pm$0.5 times more likely in RD QSOs
compared to RUD QSOs. 
\item The occurrence of multiple AAS is 2.5$\pm$0.6 times more likely in RD
QSOs than in RUD QSOs.
\item Among the RD QSOs, the frequency of occurrence of AAS appears to be
independent of the radio morphology. 
\item The frequency of occurrence of AAS depends on black hole mass, QSOs
with larger \mbh~ have higher rate of incidences. Thus, the incidence rate
is higher for older black holes. 
\item The EEFOII in the AAS samples over that in the control samples originates
at the QSO redshift, and is consistent with its origin in the QSO. The width of
the excess emission is large (FWHM $\sim$ 8-10 \aa) and is similar to the width
of the [O II] line itself. This indicates its origin in the [O II]
emitting regions in the AGN and its host galaxy. The presence of AAS enhances the O II emission from the AGN and/or the host galaxy.
\item The EEFOII is similar for RD and RUD QSOs.
\item For the RD as well as for RUD QSOs, the EEFOII depends on \wmg, such that
the sub-sample with higher \wmg~ has higher EEFOII in both cases.
\item Among the RD QSOs, the EEFOII is higher for the LD QSOs by a factor of
2.5$\pm$0.4 compared to the CD QSOs. 
\item The EEFOII depends on the mass of the black hole and the Eddington
ratio such that QSOs with higher \mbh~ and lower \edr~ have higher EEFOII.
\item The EEFOII and dust extinction in CD and LD QSOs are anticorrelated.
\item The EEFOII is similar in magnitude to that found in the intervening
absorbers.
\end{enumerate}
Based on these results the AAS seem to have very different amount of dust and
dust-to-gas-ratio as compared to the intervening systems, which seem to depend
on the radio properties of the QSOs and also on the masses of the central black
hole. The excess [O~II] emission occurs at the QSO redshift. The width of the
excess emission is similar to that of the emission lines in control samples.
The AAS could therefore be intrinsic to the QSOs.  Even with this large sample
of AAS, we are possibly able to argue against only two of the possibilities,
(i) and (iv), mentioned in section 1 for the origin of these systems.
Their higher dust content compared to the intervening systems (even for \bet
$>$ 0.005 systems among the AAS) and its dependence on QSO properties argues
against their origin in the ISM of galaxies clustering around the QSO, or in
the ISM of the host galaxies themselves. One can argue that the jets from the
AGN can influence the ISM in the host as well as the surrounding galaxies.
However, the higher extinction is seen in both radio loud and radio quiet QSOs.
Further studies are needed to distinguish between the other two scenarios. 

\section*{Acknowledgments} PK thanks CSIR, India for the grant of Emeritus
Scientist fellowship. We would like to thank Shen and Menard for making
the absorber sample as well as the control sample available. We are grateful to
the referee for giving detailed comments and suggestions which helped improve
the presentation in a significant way.


\begin{thebibliography}{}
\bibitem[Aldcroft, et al. (1993)]{Al93} Aldcroft, T. L., Bechtold, J. \& Elvis, M. 1994, ApJS, 93, 1
\bibitem[Anderson et al. (1987)]{An87} Anderson, S. F., Weymann, R. J., Foltz, C. B. \& Chaffee, F. H. 1987, AJ, 94, 278 
\bibitem[Baker et al. (2002)]{Ba02} Baker, J. C., Hunstead, R. W., Athreya, R. M., Barthel, P. D., de Silva, E., Lehnert, M. D., \& Suaners, R. D. E.  2002, ApJ, 568, 592
\bibitem[Becker et al. (1995)]{BWH95} Becker, R. H., White, R. L., \& Helfand, D. J. 1995, ApJ, 450, 559
\bibitem[Barlow\& Sargent (1997)]{BS97} Barlow, T. A., \& Sargent, W. L. W. 1997, AJ, 113, 136
\bibitem[Chelouche et al. (2008)]{Ch07} Chelouche, D., Menard, B., Bowen, D. V., \& Gnat, O. 2008, ApJ, 683, 55
\bibitem[Foltz et al. (1988)]{Fo88} Foltz, C. B., Chaffee, F. H. Jr., Weyman, R. J., Anderson, S. F.
1988. In QSO Absorption lines: Probing the Universe: Proceedings of the QSO
Absorption line meeting, Baltimore, MD, May 19-21, 1987. Cambridge and New
York, Cambridge University Press, pp53-65
\bibitem[Fu \& Stockton (2007)]{FS07} Fu, H., \& Stockton, A. 2007, ApJ, 666, 794
\bibitem[Ganguly et al. (2001)]{Gan01} Ganguly, R., Bond, N. A., Charlton, J. C., Eracleous, M., Brandt, W. N. \& Churchill, C. W. 2001, ApJ, 549, 133
\bibitem[Hewett \& Wild (2010)]{HW10} Hewett, P. C. \& Wild, V. 2010, MNRAS, 405, 2302
\bibitem[Hopkins et al. (2005)]{H05} Hopkins, P. F., Hernquist, L., Cox, T. J., et al. 2005, ApJ, 630, 705
\bibitem[Hopkins et al. (2006)]{H06} Hopkins, P. F., Hernquist, L., Cox, T. J., et al. 2006, ApJS, 163, 1
\bibitem[Khare et al. (2007)]{kh07} Khare, P., Kulkarni, V. P., Peroux, C. et al. 2007, A\&A, 464, 487
\bibitem[Kimball et al. (2011)]{K11} Kimball, A. E., Ivezic, Z., Witta, P. J. \& Schneider, D. P. 2011, AJ, 141, 182 (K11)
\bibitem[Menard et al. (2011)]{M11} Menard, B., Wild, V., Nestor, D., Quidder, A., Zibetti, S., Rao, S. \& Turnshek, D. 2011, MNRAS, 417, 801
\bibitem[Noterdaeme et al. (2010)]{N10} Noterdaeme, P. Srianand, R. \& Mohan, V. 2010, MNRAS, 403, 906
\bibitem[Pei (1992)]{P92} Pei Y. C., 1992, ApJ, 395, 130
\bibitem[Sanders et al. (1996)]{SM96} Sanders, D. B., Soifer, B. T., Elias, J. H., et al. 1988, ApJ, 325, 74
\bibitem[Shen et al. (2011)]{SH11} Shen, Y.,  Richards, G. T., Strauss, M. A., et al. 2011, ApJS, 194, 45
\bibitem[Shen \& Menard (2012)]{SM12} Shen, Y. \& Menard, B. 2012, ApJ, 748, 131 (SM12)
\bibitem[Stoughton et al. (2002)]{St02} Stoughton C., Lupton, R. H.; Bernardi, M.  et al., 2002, AJ, 123, 485.
\bibitem[Vanden Berk et al. (2008)]{V08} Vanden Berk, D. E., Khare, P., York, D. G, et al. 2008, ApJ, 679, 239 (V08) 
\bibitem[Vestergaard (2003)]{Ve03} Vestergaard, M. 2003, ApJ, 599, 116 
\bibitem[Wild et al. (2006)]{W06} Wild, V., Hewett, P. C. \& Pettini, M. 2006, MNRAS, 367, 211 
\bibitem[Wild et al. (2008)]{W08} Wild, V., Kauffmann, G., White, S., et al. 2008, MNRAS, 388, 227
\bibitem[York et al. (2006)]{Y06} York, D. G., Khare, P., Vanden Berk, D. E., et al. 2006, MNRAS, 367, 945 (Y06)
\end{thebibliography}
\end{document}